\documentclass[showpacs,aps,prd]{revtex4}
\input epsf

\begin{document}

\title{ Mass spectra of the heavy baryons $\Lambda_{Q}$ and $\Sigma_{Q}^{(*)}$ from QCD sum rules}
\author{Jian-Rong Zhang and Ming-Qiu Huang}
\affiliation{Department of Physics, National University of Defense
Technology, Hunan 410073, China}
\date{\today}
%%%%%%%%%%%%%%%%%%%%%%%%%%%%%%%%%%%%%%%%%%%%%%%%%%%%%%%%%%%%%%%%%%%%%
\begin{abstract}
We use QCD sum rule approach to calculate the masses of the
ground-state $\Lambda_{Q}$ and $\Sigma_{Q}^{(*)}$ baryons.
Contributions of the operators up to dimension six are included in
operator product expansion. The resulting heavy baryonic masses from
the calculations are $m_{\Lambda_{b}}=5.69\pm0.13~\mbox{GeV}$, and
$m_{\Lambda_{c}}=2.31\pm0.19~\mbox{GeV}$ for $\Lambda_{Q}$;
$m_{\Sigma_{b}}=5.73\pm0.21~\mbox{GeV}$,
$m_{\Sigma_{b}^{*}}=5.81\pm0.19~\mbox{GeV}$,
$m_{\Sigma_{c}}=2.40\pm0.31~\mbox{GeV}$ and
$m_{\Sigma_{c}^{*}}=2.56\pm0.24~\mbox{GeV}$ for $\Sigma_{Q}^{(*)}$,
respectively, which are in good agreement with  the experimental
values.
\end{abstract}
\maketitle
%%%%%%%%%%%%%%%%%%%%%%%%%%%%%%%%%%%%%%%%%%%%%%%%%%%%%%%%%%%%%%%%%%%%%
\section{Introduction}\label{sec1}
Recently, the CDF Collaboration has reported the observations of the
heavy baryons $\Sigma_{b}$ and $\Sigma_{b}^{*}$ \cite{sigma-b}. Up
to now, the masses of the ground-state $\Lambda_{Q}$ and
$\Sigma_{Q}^{(*)}$ $(Q=b,c)$ baryons as well as masses of several
lower lying excited heavy baryons have been measured
\cite{splitting,lamdarc,lamdar-b}. With the accumulation of
experimental data, reliable and comprehensive theoretical
explanations are needed. There were some theoretical investigations
on the heavy baryon masses, such as quark models \cite{quark
model,quark model 2}, mass formulas \cite{mass formular}, and
lattice QCD calculations \cite{lattice}.  With QCD sum rules
\cite{svzsum}, heavy baryons were first discussed in heavy-quark
limit in Ref. \cite{evs}, then masses for heavy baryons were
calculated in the heavy quark effective theory (HQET) to leading and
next-to-leading order in $\alpha_{s}$ \cite{alfa}, and to order
$1/m_{Q}$ \cite{dai,mqhuang}. In Refs. \cite{EBagan}, the
calculations for the heavy baryons began with the full theory and
results of the calculations were expanded by heavy-quark masses.
Lately, the masses of $\Xi_{Q}$ and $\Omega_{Q}^{*}$ have been
studied in QCD sum rules \cite{ksi,zhigang}. Along with the
significant observation of $\Sigma_{b}$ and $\Sigma_{b}^{*}$,
several renewed theoretical studies have been done by using various
approaches mentioned above \cite{new}. In this paper we shall study
heavy baryonic two-point correlators and obtain mass sum rules for
$\Lambda_{Q}$ and $\Sigma_{Q}^{(*)}$, using the technique developed
in \cite{kim,bracco}. The paper is organized as follows. In Sec
\ref{sec2} we have derived QCD sum rules for $\Lambda_{Q}$ and
$\Sigma_{Q}^{(*)}$. Section \ref{sec3} is devoted to numerical
analysis and discussions. This section also includes a brief
summary.
%%%%%%%%%%%%%%%%%%%%%%%%%%%%%%%%%%%%%%%%%%%%%%%%%%%%%%%%%%%%%%%%%%%%%
\section{ QCD sum rules for $\Lambda_{Q}$ and $\Sigma_{Q}^{(*)}$}\label{sec2}
   The basic points in the application of QCD sum rules to problems
involving heavy baryons are to choose certain suitable interpolating
currents in terms of quark fields. The currents for $\Lambda_{Q}$
and doublet $\{\Sigma_{Q}, \Sigma_{Q}^{*}\}$ are associated with the
spin-parity quantum numbers $j^{p}=0^{+}$ and $j^{p}=1^{+}$ for the
light diquark system with antisymmetric and symmetric flavor
structure, respectively. Adding the heavy quark to the light-quark
system, one obtains $j^{p}=1/2^{+}$ for the baryons $\Lambda_{Q}$
and the pair of degenerate states $j^{p}=1/2^{+}$ and
$j^{p}=3/2^{+}$ for the baryons $\Sigma_{Q}$ and $\Sigma_{Q}^{*}$.
Consequently, we need an isospin-0 quark pair for $\Lambda_{Q}$ and
an isospin-1 quark pair for $\Sigma_{Q}$, after adding a third quark
to form antisymmetric and symmetric flavor structure respectively,
which can determine the choice of $\gamma$ matrices in baryonic
currents \cite{evs}. For $\Sigma_{Q}^{*}$, the currents can be
obtained from those of $\Sigma_{Q}$ using SU(3) symmetry relations
\cite{Ioffe}. Concretely, we adopt the following forms of currents
for the heavy baryons $\Lambda_{Q}$ and $\Sigma_{Q}^{(*)}$, which
can be generally written as \cite{evs,Ioffe,cohen}:
\begin{eqnarray}
j_{Q}=\varepsilon_{abc}(q_{1}^{Ta}C\Gamma_{k}q_{2}^{b})\Gamma_{k}^{'}Q^{c}
\end{eqnarray}
for $j^{p}=1/2^{+}$, and
\begin{eqnarray}
j_{Q}=\varepsilon_{abc}[2/\sqrt{3}(q_{1}^{Ta}C\Gamma_{k}Q^{b})\Gamma_{k}^{'}q_{2}^{c}
+1/\sqrt{3}(q_{1}^{Ta}C\Gamma_{k}q_{2}^{b})\Gamma_{k}^{'}Q^{c}]
\end{eqnarray}
for $j^{p}=3/2^{+}$. $\Gamma_{k}$ and $\Gamma_{k}^{'}$ are chosen
covariantly as
\begin{eqnarray}
\Gamma_{k}=\gamma_{5}, \Gamma_{k}^{'}=1
\end{eqnarray}
for $\Lambda_{Q}$ baryons, and
\begin{eqnarray}
\Gamma_{k}=\gamma_{\mu}, \Gamma_{k}^{'}=\gamma_{\mu}\gamma_{5}
\end{eqnarray}
for $\Sigma_{Q}^{(*)}$ baryons. Here the index $T$ means matrix
transposition, $C$ is the charge conjugation matrix, and $a$, $b$,
$c$ are color indices. The QCD sum rules for $\Lambda_{Q}$ and
$\Sigma_{Q}^{(*)}$ are constructed from the two-point correlation
function
\begin{eqnarray}\label{correlator}
\Pi(q^{2})=i\int
d^{4}x\mbox{e}^{iq.x}\langle0|T[j_{Q}(x)\overline{j}_{Q}(0)]|0\rangle.
\end{eqnarray}
Lorentz covariance, parity, and time reversal imply that the
two-point correlation function in Eq. (\ref{correlator}) has the
form
\begin{eqnarray}
\Pi(q^{2})=\Pi_{1}(q^{2})+\rlap/q\Pi_{2}(q^{2}).
\end{eqnarray}
According to the philosophy of QCD sum rules, the correlator is
evaluated in two ways. Phenomenologically, the correlator can be
expressed as a dispersion integral over a physical spectral function
\begin{eqnarray}
\Pi(q^{2})=\lambda^{2}_H\frac{\rlap/q+M_{H}}{M_{H}^{2}-q^{2}}+\frac{1}{\pi}\int_{s_{0}}
^{\infty}ds\frac{\mbox{Im}\Pi^{\mbox{phen}}(s)}{s-q^{2}}+\mbox{subtractions},
\end{eqnarray}
where $M_{H}$ denotes the mass of the heavy baryon. In the operator
product expansion (OPE) side, short-distance effects are taken care
of by Wilson coefficients, while long-distance confinement effects
are included as power corrections and parameterized in terms of
vacuum expectation values of local operators, the so-called
condensates. Hence
\begin{eqnarray}\label{ope}
\Pi_{i}(q^{2})&=&\Pi_{i}^{\mbox{pert}}(q^{2})+\Pi_{i}^{\mbox{cond}}(q^{2}),
 i=1,2.
\end{eqnarray}
We work at leading order in $\alpha_{s}$ and consider condensates up
to dimension six. To keep the heavy-quark mass finite, we use the
momentum-space expression for the heavy-quark propagator. We follow
Refs. \cite{kim,bracco} and calculate the light-quark part of the
correlation function in the coordinate space, which is then
Fourier-transformed to the momentum space in $D$ dimension. The
resulting light-quark part is combined with the heavy-quark part
before it is dimensionally regularized at $D=4$. For the heavy-quark
propagator with two and three gluons attached we use the
momentum-space expressions given in Ref. \cite{reinders}. With Eq.
(\ref{ope}), we can write the correlation function in the OPE side
in terms of a dispersion relation
\begin{eqnarray}
\Pi_{i}(q^{2})=\int_{m_{Q}^{2}}^{\infty}ds\frac{\rho_{i}(s)}{s-q^{2}}+\Pi_{i}^{\mbox{cond}}(q^{2}),
\end{eqnarray}
where the spectral density is given by the imaginary part of the
correlation function
\begin{eqnarray}
\rho_{i}(s)=\frac{1}{\pi}\mbox{Im}\Pi_{i}^{\mbox{OPE}}(s).
\end{eqnarray}
Equating the two expressions for $\Pi(q^{2})$ and assuming
quark-hadron duality yield the sum rules, from which masses of the
heavy baryons can be determined. After making a Borel transform and
transferring the continuum contribution to the OPE side, the sum
rules can be written as
\begin{eqnarray}\label{sumrule1}
\lambda^{2}_HM_{H}e^{-M_{H}^{2}/M^{2}}&=&\int_{m_{Q}^{2}}^{s_{0}}ds\rho_{1}(s)e^{-s/M^{2}}+\hat{B}\Pi_{1}^{\mbox{cond}}
\end{eqnarray}
\begin{eqnarray}\label{sumrule2}
\lambda^{2}_He^{-M_{H}^{2}/M^{2}}&=&\int_{m_{Q}^{2}}^{s_{0}}ds\rho_{2}(s)e^{-s/M^{2}}+\hat{B}\Pi_{2}^{\mbox{cond}}.
\end{eqnarray}
To eliminate the baryon coupling constant $\lambda_H$ and extract the
resonance mass $M_{H}$, we first take the derivative of Eq.
(\ref{sumrule1}) with respect to $1/M^{2}$, divide the result by itself
and deal with Eq. (\ref{sumrule2}) in the same way to get
\begin{eqnarray}\label{sum rule m}
M_{H}^{2}&=&\{\int_{m_{Q}^{2}}^{s_{0}}ds\rho_{1}(s)s
e^{-s/M^{2}}+d/d(-\frac{1}{M^{2}})\hat{B}\Pi_{1}^{\mbox{cond}}(s)\}/
\{\int_{m_{Q}^{2}}^{s_{0}}ds\rho_{1}(s)e^{-s/M^{2}}
+\hat{B}\Pi_{1}^{\mbox{cond}}(s)\}
\end{eqnarray}
\begin{eqnarray}\label{sum rule q}
M_{H}^{2}&=&\{\int_{m_{Q}^{2}}^{s_{0}}ds\rho_{2}(s)s
e^{-s/M^{2}}+d/d(-\frac{1}{M^{2}})\hat{B}\Pi_{2}^{\mbox{cond}}(s)\}/
\{\int_{m_{Q}^{2}}^{s_{0}}ds\rho_{2}(s)e^{-s/M^{2}}
+\hat{B}\Pi_{2}^{\mbox{cond}}(s)\},
\end{eqnarray}
where
\begin{eqnarray}
\rho_{i}(s)=\rho_{i}^{\mbox{pert}}(s)+\rho_{i}^{\langle\bar{q}q\rangle}(s)+\rho_{i}^{\langle
G^{2}\rangle}(s)
\end{eqnarray}
with
\begin{eqnarray}
\rho_{1}^{\mbox{pert}}(s)&=&\frac{3}{2^{7}\pi^{4}}m_{Q}\int_{\Lambda}^{1}d\alpha(\frac{1-\alpha}{\alpha})^{2}(m_{Q}^{2}-s\alpha)^{2}\\
\rho_{1}^{\langle G^{2}\rangle}(s)&=&\frac{\langle
g^{2}G^{2}\rangle}{2^{9}\pi^{4}}m_{Q}\int_{\Lambda}^{1}d\alpha[(\frac{1-\alpha}{\alpha})^{2}+2]\\
\hat{B}\Pi_{1}^{\mbox{cond}}&=&-\frac{\langle
g^{2}G^{2}\rangle}{3\cdot2^{9}\pi^{4}}m_{Q}^{3}\int_{0}^{1}d\alpha\frac{(1-\alpha)^{2}}{\alpha^{3}}e^{-m_{Q}^{2}/(\alpha M^{2})}\nonumber\\
& &{}+\frac{\langle\bar{q}q\rangle^{2}}{6}m_{Q}e^{-m_{Q}^{2}/M^{2}}\nonumber\\
& & {}-\frac{\langle
g^{3}G^{3}\rangle}{3\cdot2^{10}\pi^{4}}m_{Q}\int_{0}^{1}
d\alpha\frac{(1-\alpha)^{2}}{\alpha^{3}}(3-\frac{m_{Q}^{2}}{\alpha
M^{2}})e^{-m_{Q}^{2}/(\alpha M^{2})}\\
\rho_{2}^{\mbox{pert}}(s)&=&\frac{3}{2^{7}\pi^{4}}\int_{\Lambda}^{1}d\alpha\frac{(1-\alpha)^{2}}{\alpha}(m_{Q}^{2}-s\alpha)^{2}\\
\rho_{2}^{\langle G^{2}\rangle}(s)&=&\frac{\langle
g^{2}G^{2}\rangle}{2^{9}\pi^{4}}[1-(\frac{m_{Q}^{2}}{s})^{2}]\\
\hat{B}\Pi_{2}^{\mbox{cond}}&=&-\frac{\langle
g^{2}G^{2}\rangle}{3\cdot2^{9}\pi^{4}}m_{Q}^{2}\int_{0}^{1}d\alpha(\frac{1-\alpha}{\alpha})^{2}e^{-m_{Q}^{2}/(\alpha M^{2})}\nonumber\\
& &{}+\frac{\langle\bar{q}q\rangle^{2}}{6}e^{-m_{Q}^{2}/M^{2}}\nonumber\\
& & {}-\frac{\langle
g^{3}G^{3}\rangle}{3\cdot2^{11}\pi^{4}}\int_{0}^{1}
d\alpha(\frac{1-\alpha}{\alpha})^{2}(1-\frac{2m_{Q}^{2}}{\alpha
M^{2}})e^{-m_{Q}^{2}/(\alpha M^{2})}
\end{eqnarray}
for $\Lambda_{Q}$ baryons,
\begin{eqnarray}
\rho_{1}^{\mbox{pert}}(s)&=&\frac{3}{2^{4}\pi^{4}}m_{Q}\int^{1}_{\Lambda}d\alpha(\frac{1-\alpha}{\alpha})^{2}(m_{Q}^{2}-s\alpha)^{2}\\
\rho_{1}^{\langle G^{2}\rangle}(s)&=&\frac{\langle
g^{2}G^{2}\rangle}{2^{6}\pi^{4}}m_{Q}\int_{\Lambda}^{1}d\alpha[(\frac{1-\alpha}{\alpha})^{2}-2]\\
\hat{B}\Pi_{1}^{\mbox{cond}}&=&
-\frac{\langle g^{2}G^{2}\rangle}{3\cdot2^{6}\pi^{4}}m_{Q}^{3}\int^{1}_{0}d\alpha\frac{(1-\alpha)^{2}}{\alpha^{3}}e^{-m_{Q}^{2}/(\alpha M^{2})}\nonumber\\
& & {}+\frac{8\langle\bar{q}q\rangle^{2}}{3}m_{Q}e^{-m_{Q}^{2}/M^{2}}\nonumber\\
& & {}-\frac{\langle g^{3}G^{3}\rangle}{3\cdot2^{7}\pi^{4}}m_{Q}
\int_{0}^{1}
d\alpha\frac{(1-\alpha)^{2}}{\alpha^{3}}(3-\frac{m_{Q}^{2}}{\alpha
M^{2}})e^{-m_{Q}^{2}/(\alpha M^{2})}\\
\rho_{2}^{\mbox{pert}}(s)&=&\frac{3}{2^{5}\pi^{4}}\int^{1}_{\Lambda}d\alpha\frac{(1-\alpha)^{2}}{\alpha}(m_{Q}^{2}-s\alpha)^{2}\\
\rho_{2}^{\langle G^{2}\rangle}(s)&=&-\frac{\langle
g^{2}G^{2}\rangle}{2^{7}\pi^{4}}[1-(m_{Q}^{2}/s)^{2}]\\
\hat{B}\Pi_{2}^{\mbox{cond}}&=&
-\frac{\langle g^{2}G^{2}\rangle}{3\cdot2^{7}\pi^{4}}m_{Q}^{2}  \int^{1}_{0}d\alpha(\frac{1-\alpha}{\alpha})^{2}e^{-m_{Q}^{2}/(\alpha M^{2})}\nonumber\\
& & {}+\frac{4\langle\bar{q}q\rangle^{2}}{3}e^{-m_{Q}^{2}/M^{2}}\nonumber\\
& & {}-\frac{\langle g^{3}G^{3}\rangle}{3\cdot2^{9}\pi^{4}}
\int_{0}^{1}
d\alpha(\frac{1-\alpha}{\alpha})^{2}(1-\frac{2m_{Q}^{2}}{\alpha
M^{2}})e^{-m_{Q}^{2}/(\alpha M^{2})}
\end{eqnarray}
for $\Sigma_{Q}$ baryons, and
\begin{eqnarray}
\rho_{1}^{\mbox{pert}}(s)&=&\frac{1}{2^{4}\pi^{4}}m_{Q}\int^{1}_{\Lambda}d\alpha(\frac{1-\alpha}{\alpha})^{2}(m_{Q}^{2}-s\alpha)^{2}\\
\rho_{1}^{\langle\bar{q}q\rangle}(s)&=&\frac{4}{3\pi^{2}}\langle\bar{q}q\rangle
\int_{\Lambda}^{1}d\alpha(m_{Q}^{2}-s\alpha)\\
\rho_{1}^{\langle G^{2}\rangle}(s)&=&\frac{\langle
g^{2}G^{2}\rangle}{3\cdot2^{6}\pi^{4}}m_{Q}\int_{\Lambda}^{1}d\alpha[(\frac{1-\alpha}{\alpha})^{2}-2]\\
\hat{B}\Pi_{1}^{\mbox{cond}}&=&
-\frac{\langle g^{2}G^{2}\rangle}{3^{2}\cdot2^{6}\pi^{4}}m_{Q}^{3}\int^{1}_{0}d\alpha\frac{(1-\alpha)^{2}}{\alpha^{3}}e^{-m_{Q}^{2}/(\alpha M^{2})}\nonumber\\
& & {}+\frac{5\cdot2^{3}\langle\bar{q}q\rangle^{2}}{3^{2}}m_{Q}e^{-m_{Q}^{2}/M^{2}}\nonumber\\
& & {}-\frac{\langle g^{3}G^{3}\rangle}{3^{2}\cdot2^{7}\pi^{4}}m_{Q}
\int_{0}^{1}
d\alpha\frac{(1-\alpha)^{2}}{\alpha^{3}}(3-\frac{m_{Q}^{2}}{\alpha
M^{2}})e^{-m_{Q}^{2}/(\alpha M^{2})}\\
\rho_{2}^{\mbox{pert}}(s)&=&\frac{1}{2^{5}\pi^{4}}\int_{\Lambda}^{1}d\alpha\frac{(1-\alpha)^{2}(-1+3\alpha)}{\alpha^{2}}(m_{Q}^{2}-s\alpha)^{2}\\
\rho_{2}^{\langle\bar{q}q\rangle}(s)&=&-\frac{2}{3\pi^{2}}\langle\bar{q}q\rangle
m_{Q}(1-m_{Q}^{2}/s)^{2}\\
\rho_{2}^{\langle G^{2}\rangle}(s)&=&\frac{\langle
g^{2}G^{2}\rangle}{3\cdot2^{6}\pi^{4}}[-\frac{1}{2}+\frac{(m_{Q}^{2}/s)^2}{2}+\int_{\Lambda}^{1}d\alpha(\alpha-1)(m_{Q}^{2}-s\alpha)]\\
\hat{B}\Pi_{2}^{\mbox{cond}}&=&\frac{\langle g^{2}G^{2}\rangle}{3^{2}\cdot2^{7}\pi^{4}}m_{Q}^{2}\int_{0}^{1}d\alpha \frac{(1-\alpha)^{2}(1-3\alpha)}{\alpha^{3}}e^{-m_{Q}^{2}/(\alpha M^{2})}\nonumber\\
& & {}+\frac{\langle g\bar{q}\sigma\cdot G
q\rangle}{3\pi^{2}}m_{Q}\int_{0}^{1}d\alpha e^{-m_{Q}^{2}/(\alpha
M^{2})}\nonumber\\
& &
{}+\frac{4\langle\bar{q}q\rangle^{2}}{3^{2}}e^{-m_{Q}^{2}/M^{2}}\nonumber\\
& & {}+\frac{\langle
g^{3}G^{3}\rangle}{3^{2}\cdot2^{9}\pi^{4}}\int_{0}^{1}
d\alpha\frac{1-\alpha}{\alpha^{4}}[\alpha(1-4\alpha-3\alpha^{2})
-2(1-4\alpha+\alpha^{2})\frac{m_{Q}^{2}}{M^{2}}]e^{-m_{Q}^{2}/(\alpha
M^{2})}
\end{eqnarray}
for $\Sigma_{Q}^{*}$ baryons. The lower limit of  integration is
given by $\Lambda=m_{Q}^{2}/s$.
%%%%%%%%%%%%%%%%%%%%%%%%%%%%%%%%%%%%%%%%%%%%%%%%%%%%%%%%%%%%%%%%%%%
\section{Numerical results and discussions}\label{sec3}
 In the numerical analysis of the sum rules obtained above, the input values
used for the quark masses and condensates are taken as:
$m_{c}=1.25\pm0.09~\mbox{GeV}, m_{b}=4.20\pm0.07~\mbox{GeV}$
\cite{PDG} with
$\langle\bar{q}q\rangle=-(0.23)^{3}~\mbox{GeV}^{3},\\\langle
g\bar{q}\sigma\cdot G q\rangle=m_{0}^{2}\langle\bar{q}q\rangle,
m_{0}^{2}=0.8~\mbox{GeV}^{2}, \langle
g^{2}G^{2}\rangle=0.5~\mbox{GeV}^{4}$, and $\langle
g^{3}G^{3}\rangle=0.045~\mbox{GeV}^{6}$ \cite{bracco}. The proper
areas of the thresholds can be determined from the consideration
that the stability of the Borel curves should not be sensitive to
them. According to the standard criterion in QCD sum rules, the
Borel windows are fixed in such a way \cite{ksi,bracco,borel}: on
one hand, in comparison with the condensate contributions, the
perturbative contribution should be larger, and the lower limit
constraint for $M^{2}$ in the sum rule windows is obtained; on the
other hand, the upper limit constraint is obtained by imposing the
restriction that the QCD continuum contribution should be smaller
than pole contribution. Giving an illustration, the comparison
between pole and continuum contributions from Eq. (12) for
$\sqrt{s_{0}}=6.5~\mbox{GeV}$ for $\Sigma_{b}$ is shown in Fig.
1(a), and its OPE convergence by comparing the different
contributions is shown in Fig. 1(b). The analysis for the others has
similarly been done, but the corresponding figures are not listed in
the paper for conciseness. Accordingly, the thresholds and Borel
windows are taken as $\sqrt{s_0}=6.2-6.4~\mbox{GeV}$,
$M^{2}=4.5-6.0~\mbox{GeV}^{2}$ for $\Lambda_{b}$,
$\sqrt{s_0}=2.7-2.9~\mbox{GeV}$, $M^{2}=1.5-3.0~\mbox{GeV}^{2}$ for
$\Lambda_{c}$, $\sqrt{s_0}=6.4-6.6~\mbox{GeV}$,
$M^{2}=4.5-6.0~\mbox{GeV}^{2}$ for $\Sigma_{b}$ and
$\Sigma_{b}^{*}$, $\sqrt{s_0}=3.0-3.2~\mbox{GeV}$,
$M^{2}=1.5-3.0~\mbox{GeV}^{2}$ for $\Sigma_{c}$, and
$\sqrt{s_0}=3.1-3.3~\mbox{GeV}$, $M^{2}=1.5-3.0~\mbox{GeV}^{2}$ for
$\Sigma_{c}^{*}$. The Borel curves for the dependence on $M^2$ of
the heavy baryon masses are shown in Figs. 2-4.

\begin{figure}
\centerline{\epsfysize=4.8truecm\epsfbox{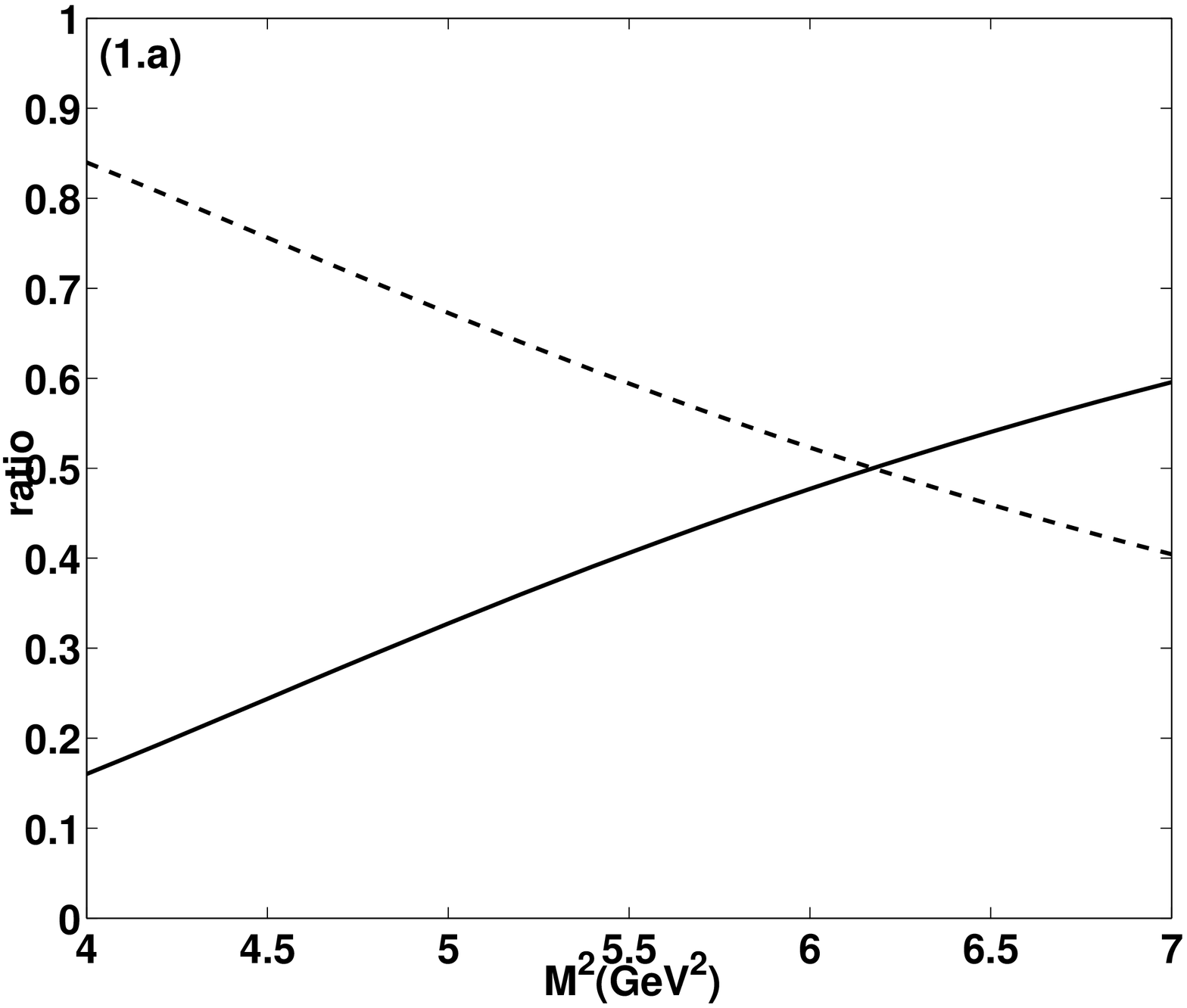}\epsfysize=4.8truecm\epsfbox{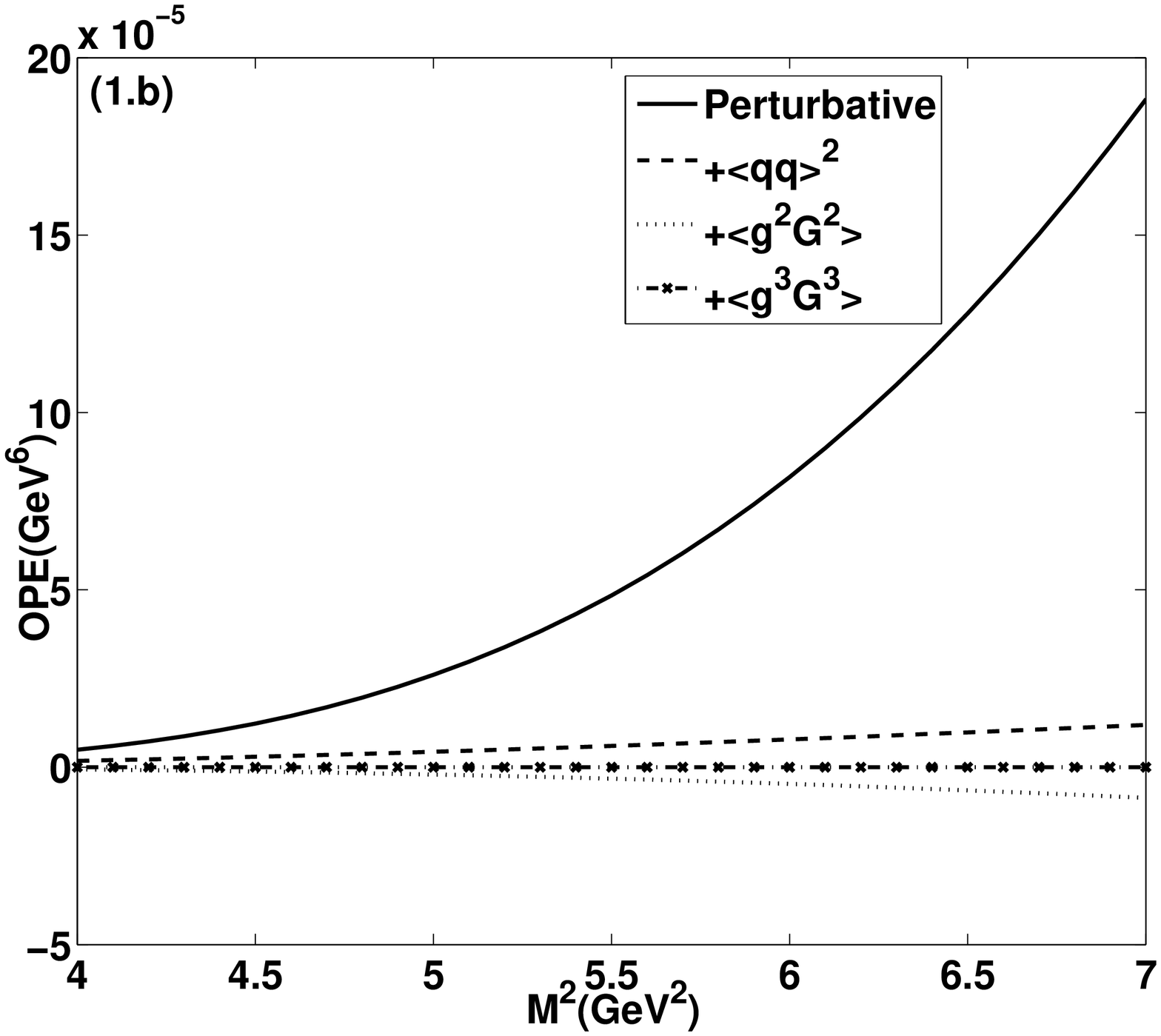}}
\caption{In (a) the dashed line shows the relative pole contribution
(the pole contribution divided by the total, pole plus continuum
contribution) and the solid line shows the relative continuum
contribution from Eq. (12) for $\sqrt{s_{0}}=6.5~\mbox{GeV}$ for
$\Sigma_{b}$. Its OPE convergence is shown in (b) by comparing the
perturbative, quark condensate, two-gluon condensate and three-gluon
condensate contributions.} \label{fig:1}
\end{figure}

\begin{figure}
\centerline{\epsfysize=4.8truecm
\epsfbox{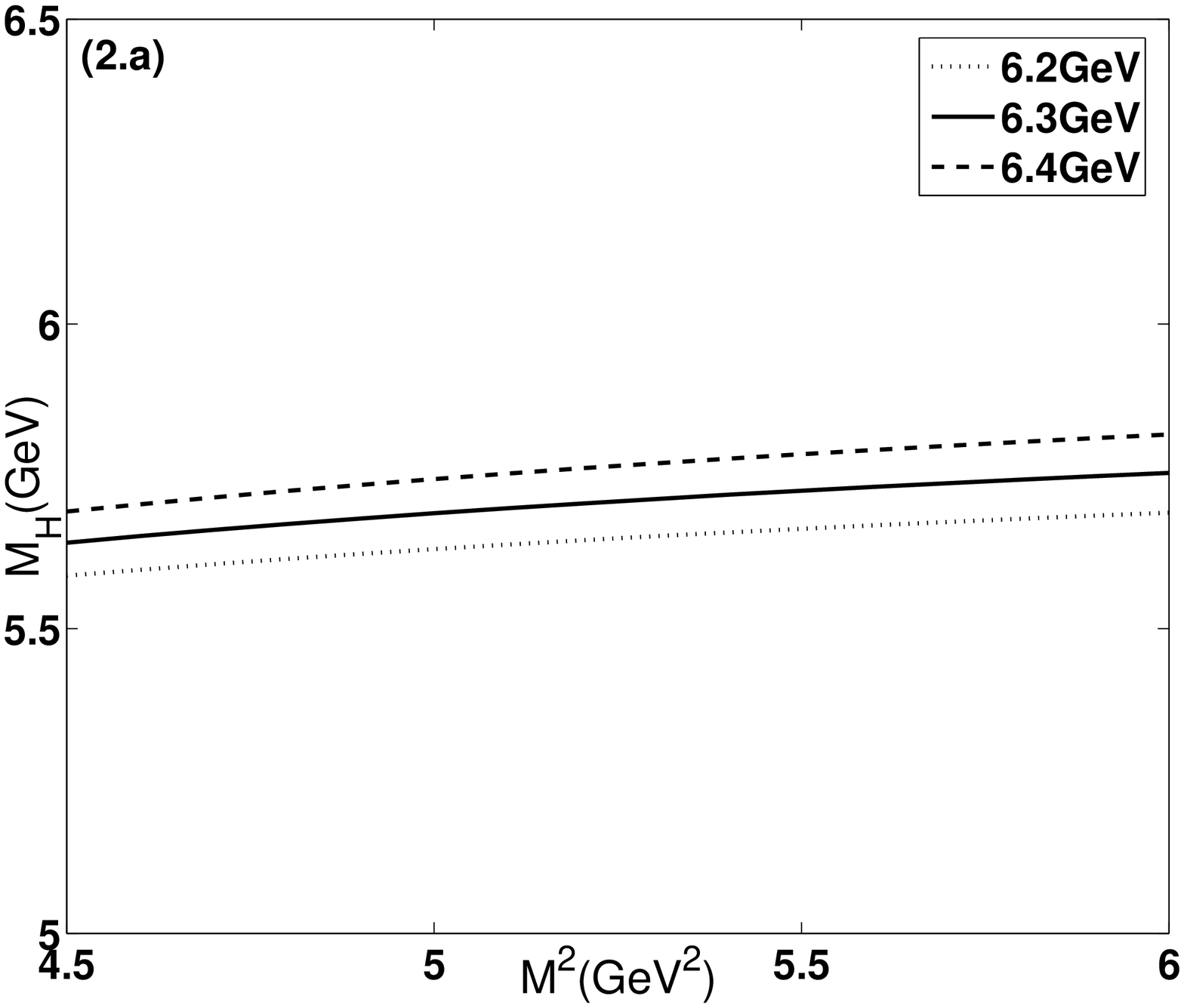}\epsfysize=4.8truecm \epsfbox{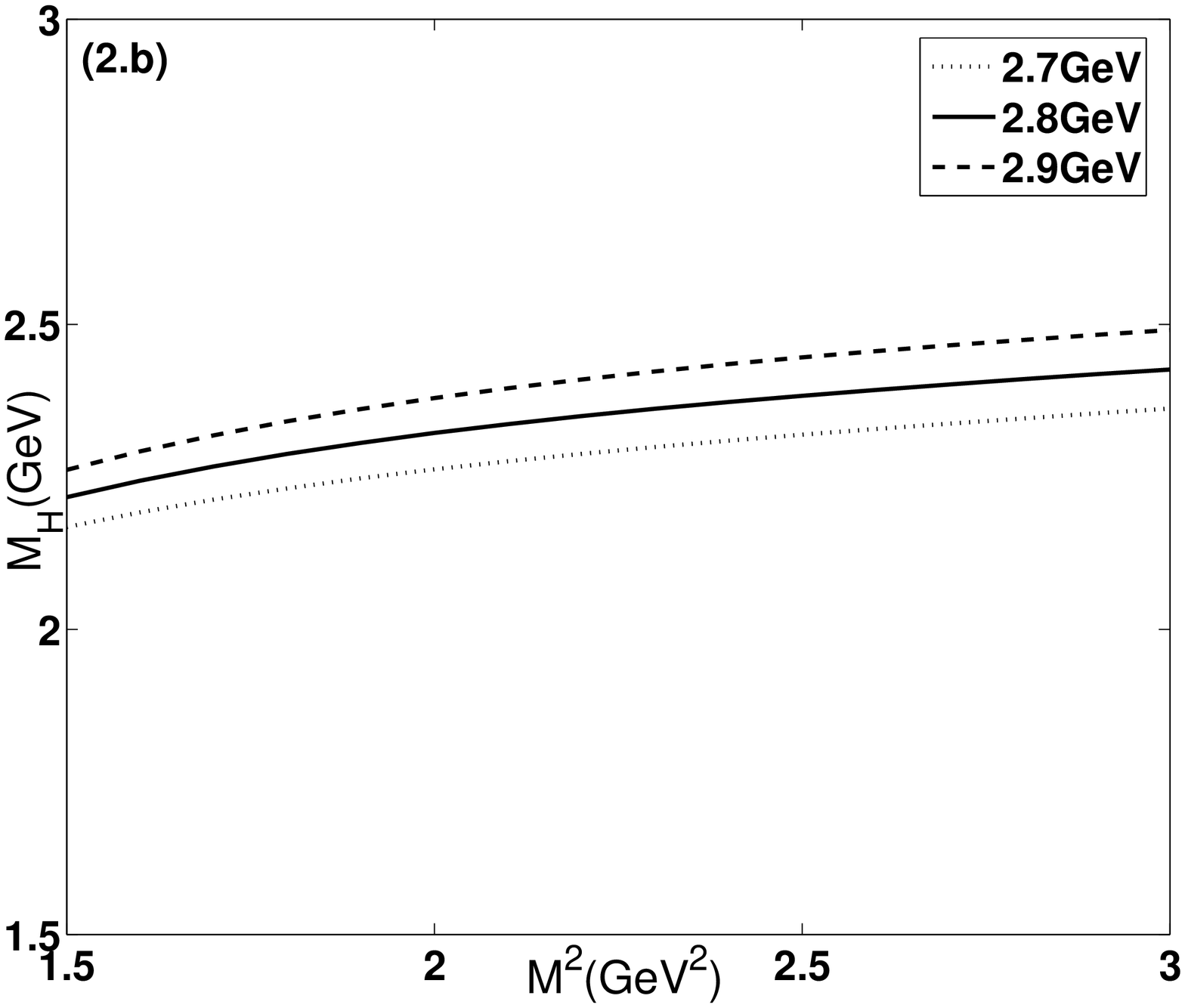}}
\centerline{\epsfysize=4.8truecm
\epsfbox{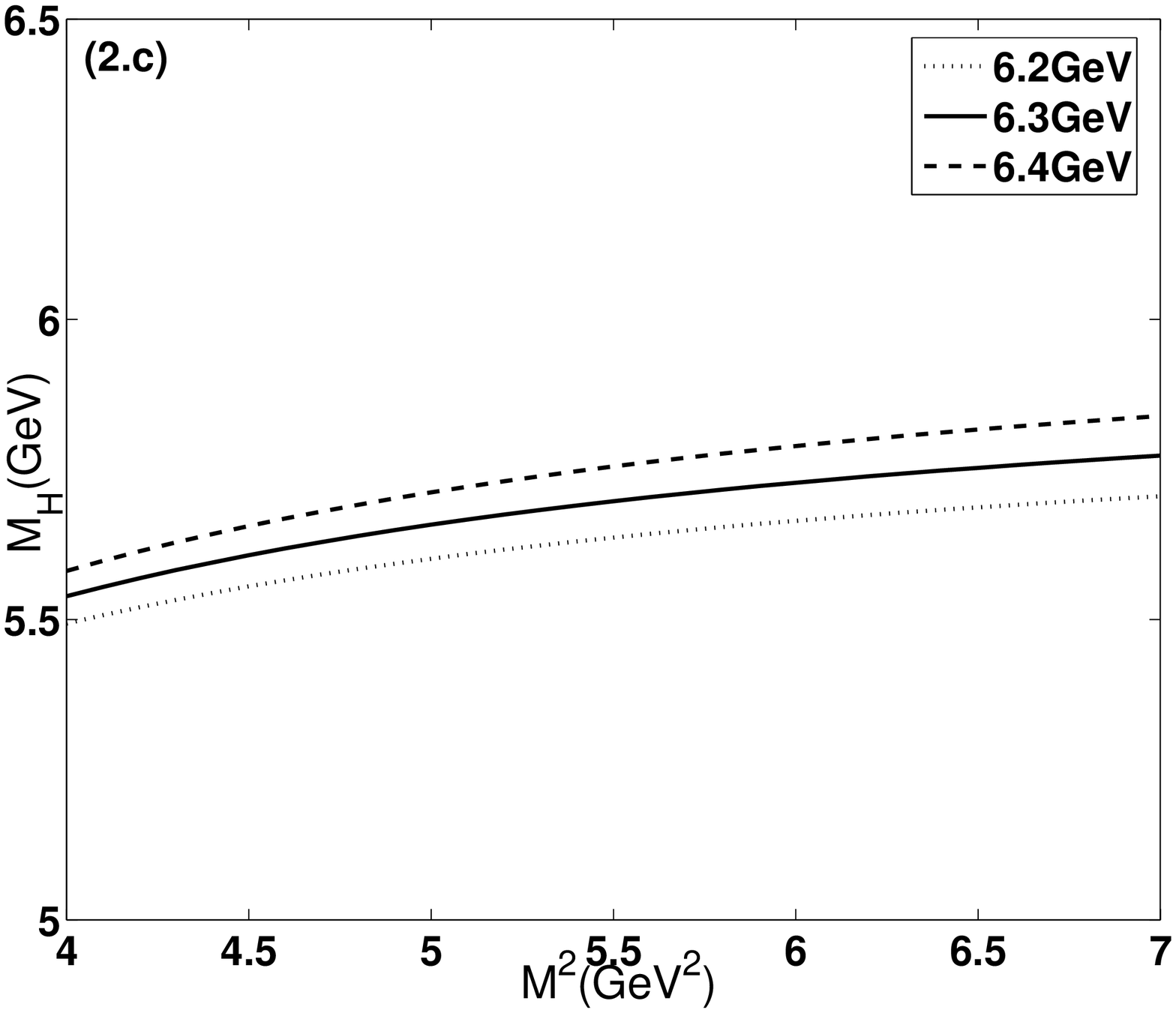}\epsfysize=4.8truecm
\epsfbox{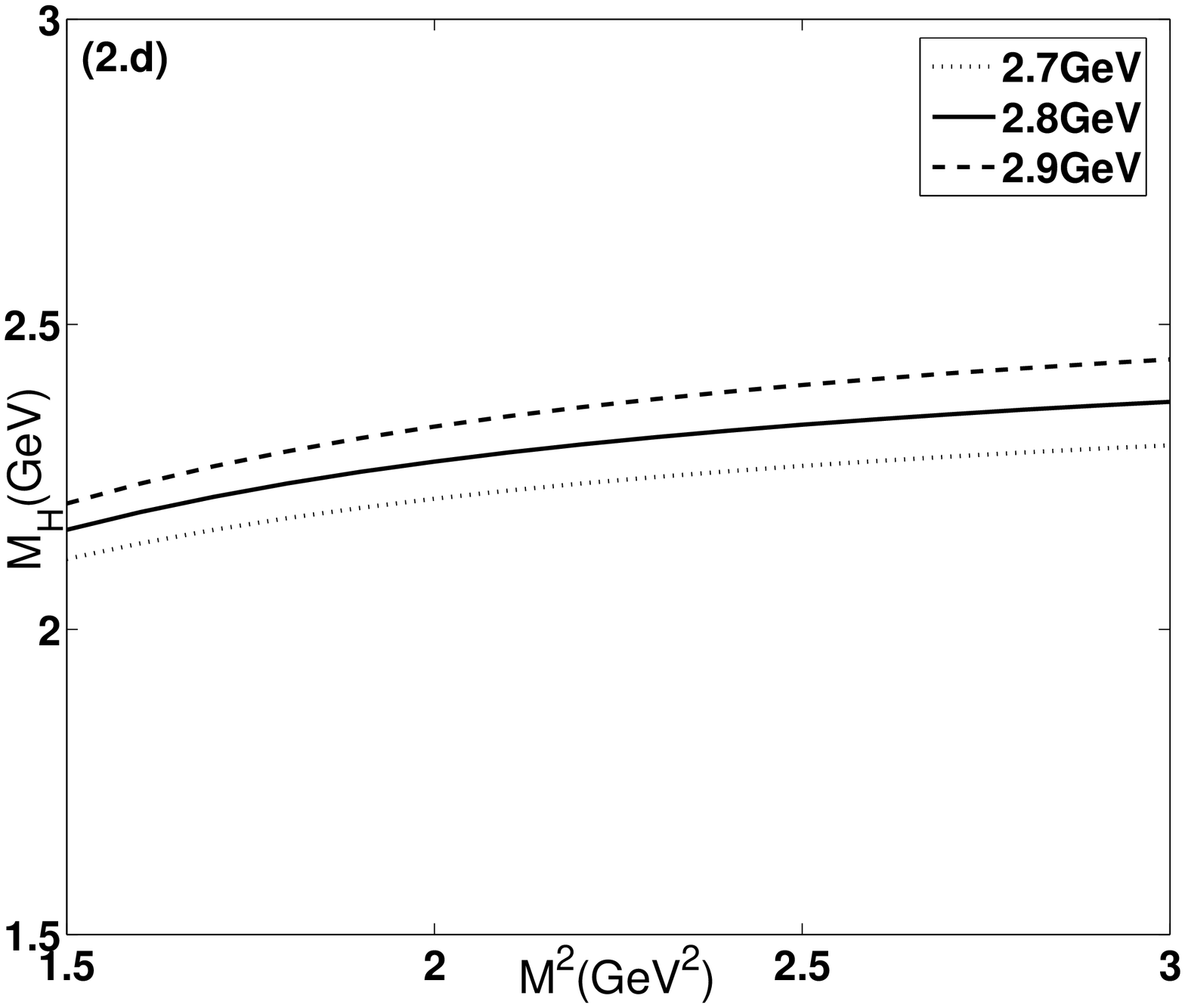}}\caption{The dependence on $M^2$ for the
masses of  $\Lambda_{b}$ and $\Lambda_{c}$. The continuum thresholds
are taken as $\sqrt{s_0}=6.2-6.4~\mbox{GeV}$,
$\sqrt{s_0}=2.7-2.9~\mbox{GeV}$ respectively: (a) and (b) are from
sum rule (\ref{sum rule m}), (c) and (d) from sum rule (\ref{sum
rule q}).} \label{fig:3}
\end{figure}

\begin{figure}
\centerline{\epsfysize=4.8truecm
\epsfbox{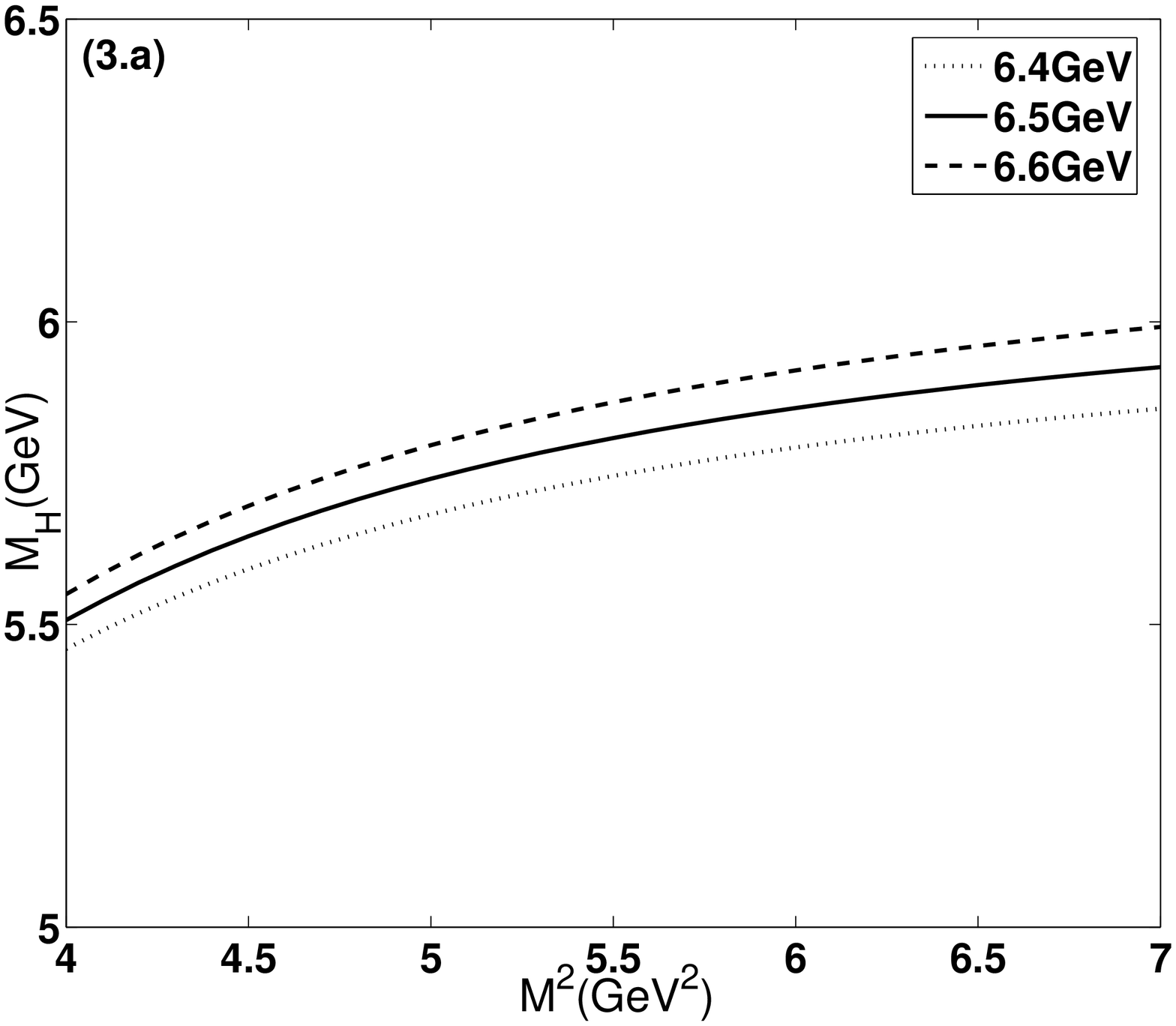}\epsfysize=4.8truecm\epsfbox{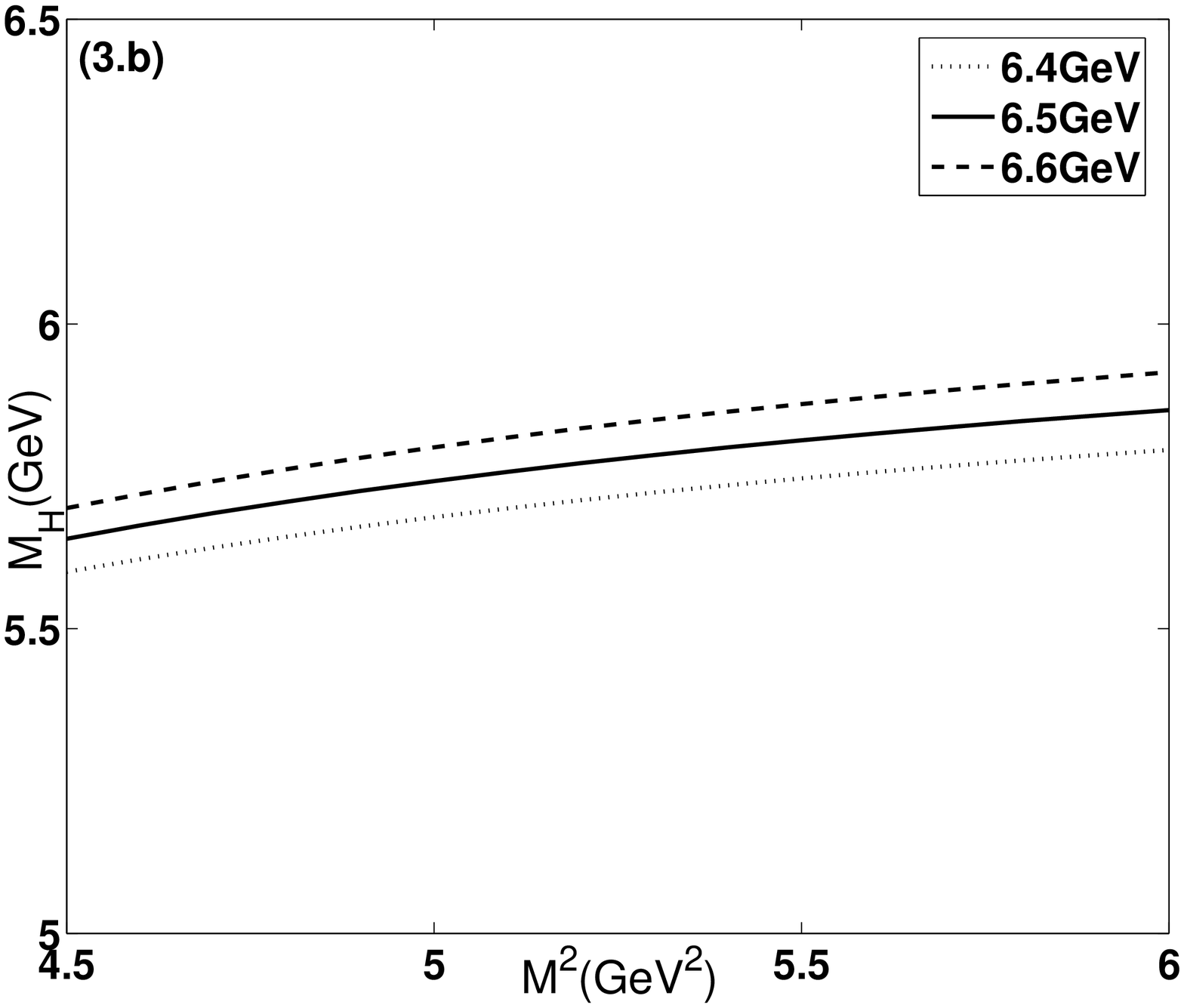}}
\centerline{\epsfysize=4.8truecm
\epsfbox{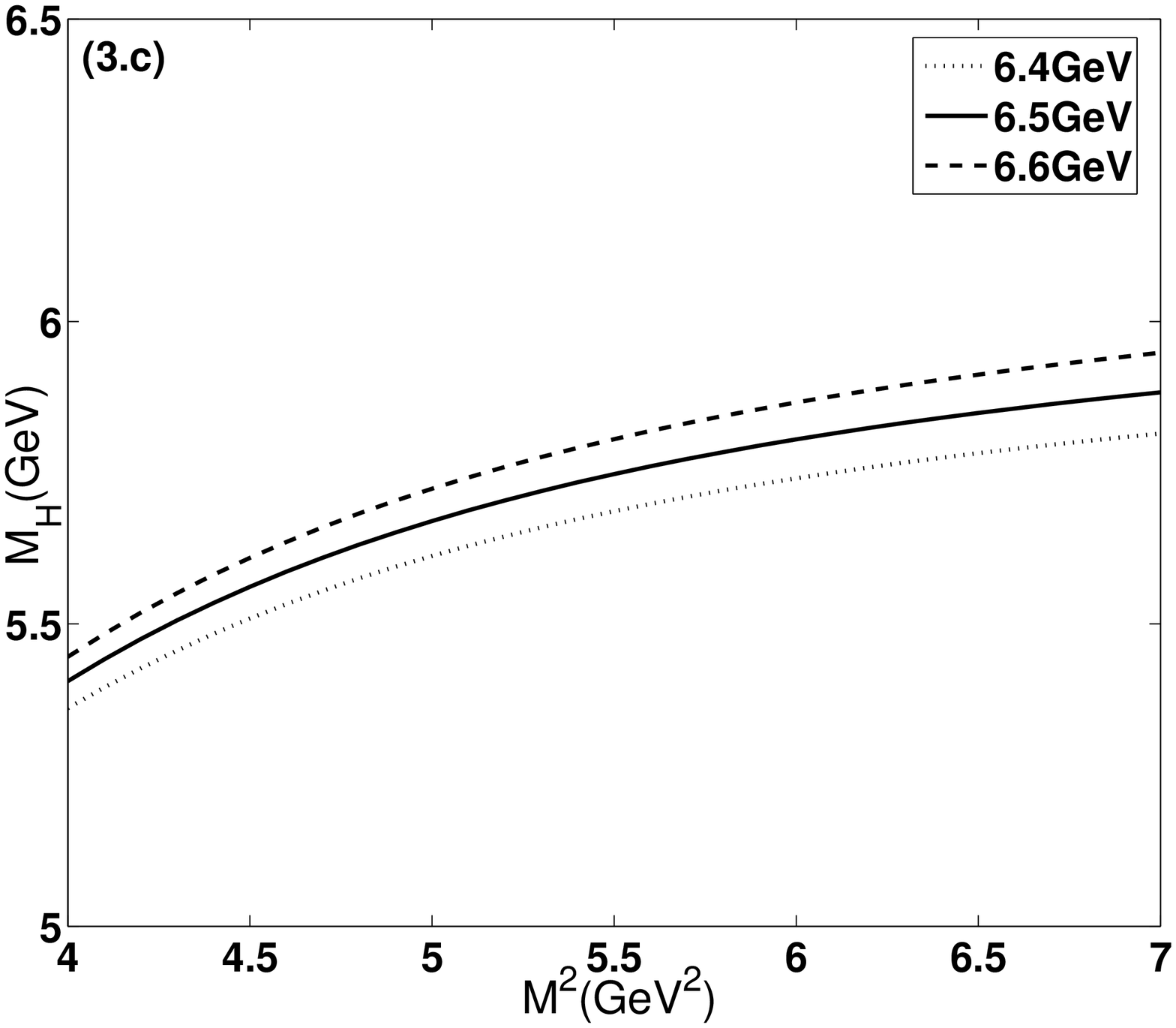}\epsfysize=4.8truecm
\epsfbox{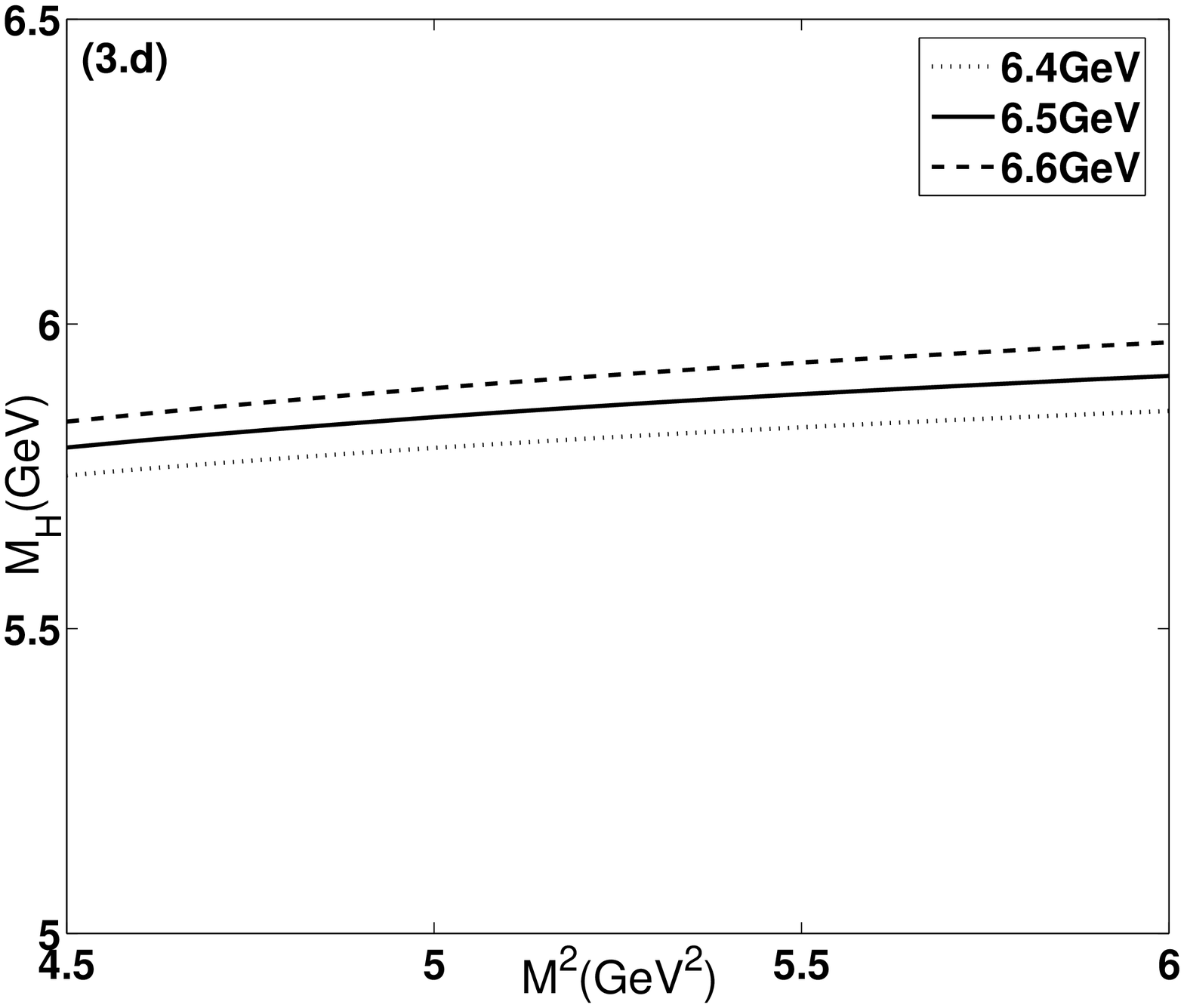}}\caption{The dependence on $M^2$ for the
masses of  $\Sigma_{b}$, $\Sigma_{b}^{*}$. The continuum thresholds
are taken as $\sqrt{s_0}=6.4-6.6~\mbox{GeV}$: (a) and (b) are from
sum rule (\ref{sum rule m}), (c) and (d) from sum rule (\ref{sum
rule q}).} \label{fig:4}
\end{figure}

\begin{figure}
\centerline{\epsfysize=4.8truecm
\epsfbox{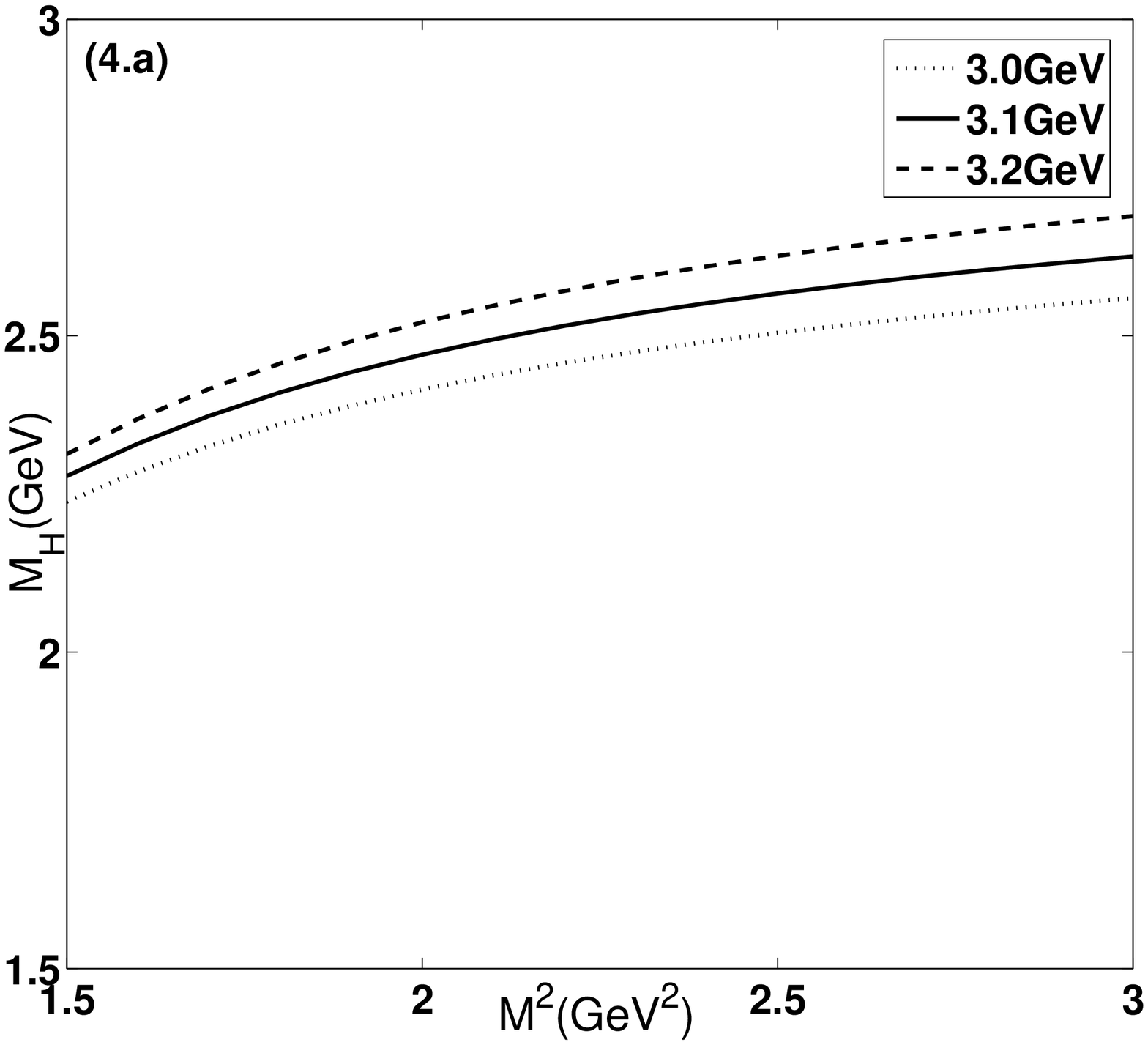}\epsfysize=4.8truecm
\epsfbox{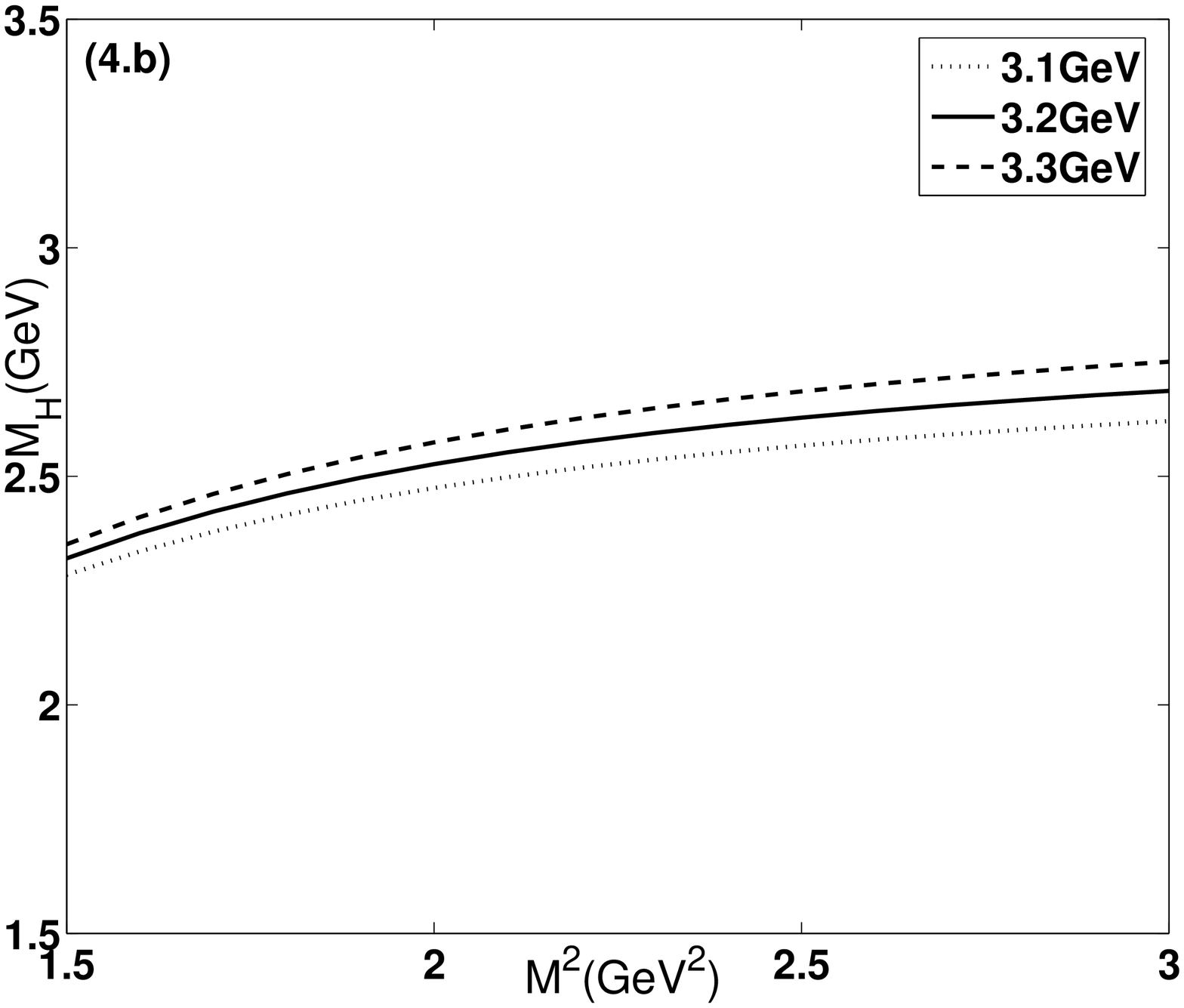}} \centerline{\epsfysize=4.8truecm
\epsfbox{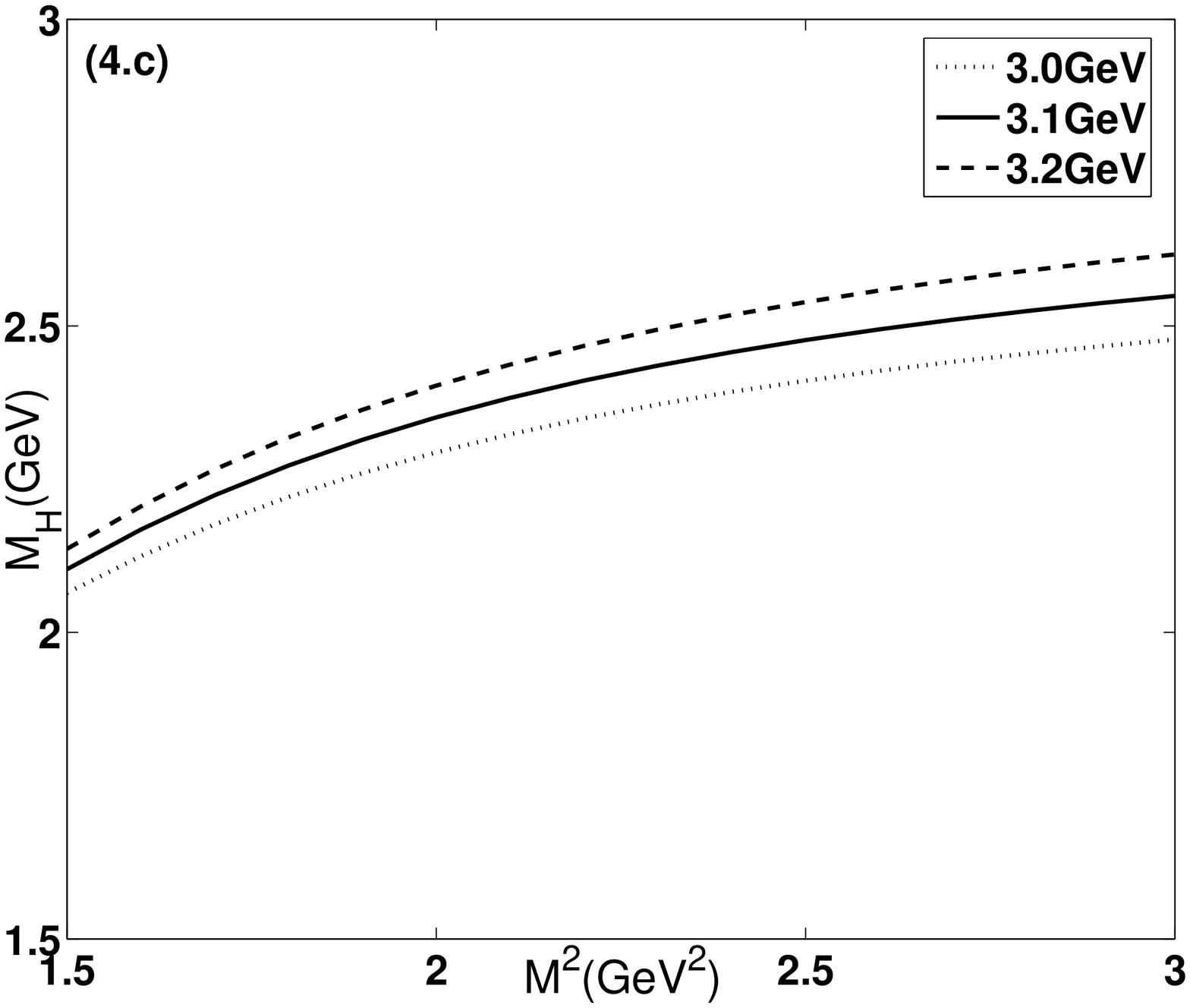}\epsfysize=4.8truecm
\epsfbox{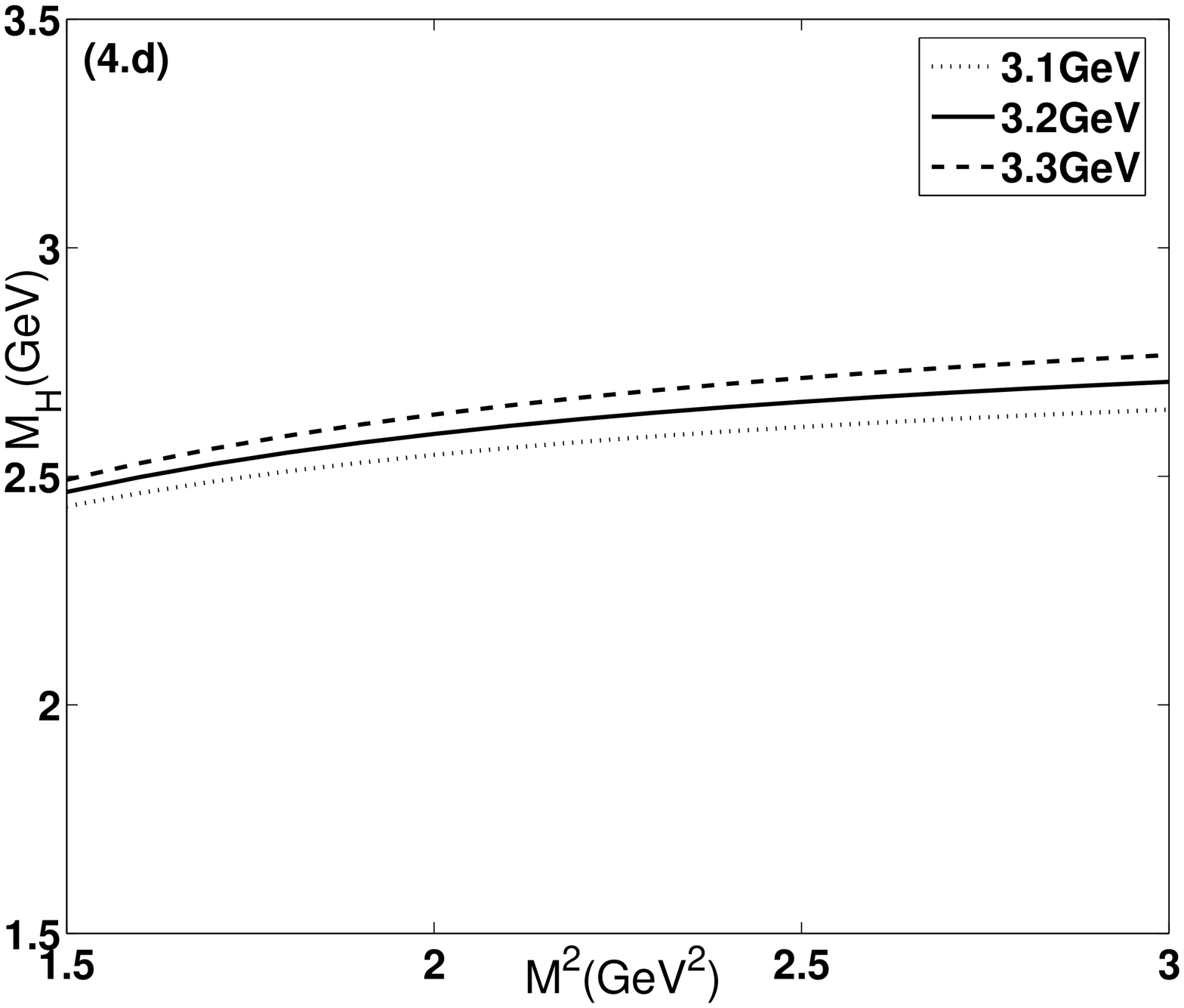}}\caption{The dependence on $M^2$ for the
masses of  $\Sigma_{c}$ and $\Sigma_{c}^{*}$. The continuum
thresholds are taken as $\sqrt{s_0}=3.0-3.2~\mbox{GeV}$,
$\sqrt{s_0}=3.1-3.3~\mbox{GeV}$ respectively: (a) and (b) are from
sum rule (\ref{sum rule m}), (c) and (d) from sum rule (\ref{sum
rule q}).} \label{fig:5}
\end{figure}

In Table \ref{table:1}, we present our results for the masses of
$\Lambda_{Q}$ and $\Sigma_{Q}^{(*)}$ baryons and a comparison  with
experimental data and other theoretical approaches. In order to
decrease the systematic errors of the sum rules we take the average
of the results obtained from sum rules (\ref{sum rule m}) and
(\ref{sum rule q}) in the numerical evaluation. The errors reflect
the uncertainty due to sum rule windows only; the uncertainty due to
the variation of the quark masses and QCD parameters is not
included. It is worth noting that the QCD $O(\alpha_s)$ corrections
in the perturbative expansion of the OPE have not been included in
the sum rule calculations. However, it is expected that the QCD
$O(\alpha_s)$ corrections might be under control since a partial
cancellation occurs in the ratio obtaining the mass sum rules
(\ref{sum rule m}) and (\ref{sum rule q}). This has been proved to
be true in the analysis for the heavy baryons in the HQET
\cite{alfa} and for the heavy mesons in full QCD
\cite{reinders,Narison}.

\begin{table}\caption{the mass spectra of $\Lambda_{Q}$ and
 $\Sigma_{Q}^{(*)}$}
\centerline{
 \begin{tabular}{ c  c  c  c  c  c  c  c c}\hline\hline
Baryon              &  Experiment~(\mbox{MeV})
      &                             &                                       &  Theory                               &                                         &                                  &                                  \\
                      &
                              & This work~(\mbox{GeV})
                              & Ref. \cite{quark model}~(\mbox{MeV})
                                & Ref. \cite{quark model 2}~(\mbox{MeV})& Ref.
                                \cite{mass formular}~(\mbox{MeV})  & Ref.
                                 \cite{lattice}~(\mbox{MeV}) &  Ref. \cite{EBagan}~(\mbox{GeV}) \\\hline
 $\Lambda_{b}$        & $5619.7\pm1.2$ \cite{lamdar-b}                                               & $5.69\pm0.13$               & 5622                                  &   5585                                &   5620                                  &    5672                          &                                  \\
                      & $5624\pm9$ \cite{PDG}                                                        &                             &                                       &                                       &                                         &                                  &                                  \\
 \hline
 $\Lambda_{c}$        & $2286.46\pm0.14$ \cite{lamdarc,PDG}                                          & $2.31\pm0.19$               & 2297                                  &   2265                                &   2285                                  &    2290                          &                                  \\
 \hline
 $\Sigma_{b}$         & $5807.8_{-2.2}^{+2.0}\pm1.7$ for $\Sigma_{b}^{+}$ \cite{sigma-b}             & $5.73\pm0.21$               & 5805                                  &   5795                                &   5820                                  &    5847                          &   $5.70\sim6.62$                 \\
                      & $5815.2\pm1.0\pm1.7$ for $\Sigma_{b}^{-}$ \cite{sigma-b}                     &                             &                                       &                                       &                                         &                                  &                                  \\
 \hline
 $\Sigma_{b}^{*}$     & $5829.0_{-1.8}^{+1.6}$$_{-1.8}^{+1.7}$ for $\Sigma_{b}^{*+}$ \cite{sigma-b}  & $5.81\pm0.19$               & 5834                                  &   5805                                &   5850                                  &    5871                          &   $5.4\sim6.2$                   \\
                      & $5836.4\pm2.0_{-1.7}^{+1.8}$ for $\Sigma_{b}^{*-}$ \cite{sigma-b}            &                             &                                       &                                       &                                         &                                  &                                  \\
 \hline
 $\Sigma_{c}$         & $2453.76\pm0.18$ \cite{PDG}                                                  & $2.40\pm0.31$               & 2439                                  &   2440                                &   2453                                  &    2452                          &   $ 2.45\sim2.94$                \\
 \hline
 $\Sigma_{c}^{*}$     & $2518.0\pm0.5$ \cite{PDG}                                                    & $2.56\pm0.24$               & 2518                                  &   2495                                &   2520                                  &    2538                          &   $2.15\sim2.92$                 \\
 \hline\hline
\end{tabular}}
\label{table:1}
\end{table}

%%%%%%%%%%%%%%%%%%%%%%%%%%%%%%%%%%%%%%%%%%%%%%%%%%%%%%%%%%%%%%%%%%%%%%%%%%%%%%%
In summary, we have applied the QCD sum rule approach to calculate
the masses of the heavy baryons $\Lambda_{Q}$ and $\Sigma_{Q}^{(*)}$
including the contributions of the operators up to dimension six in
OPE. The final results extracted from our sum rules are:
$m_{\Lambda_{b}}=5.69\pm0.13~\mbox{GeV}$,
$m_{\Lambda_{c}}=2.31\pm0.19~\mbox{GeV}$,
$m_{\Sigma_{b}}=5.73\pm0.21~\mbox{GeV}$,
$m_{\Sigma_{b}^{*}}=5.81\pm0.19~\mbox{GeV}$,
$m_{\Sigma_{c}}=2.40\pm0.31~\mbox{GeV}$, and
$m_{\Sigma_{c}^{*}}=2.56\pm0.24~\mbox{GeV}$. The gained masses are
well compatible with recent experimental data. However, there are
still some differences from our central values and experimental
values, which implies that the predictions might be improved by the
computation of the QCD $O(\alpha_s)$ corrections.

%%%%%%%%%%%%%%%%%%%%%%%%%%%%%%%%%%%%%%
\begin{acknowledgments}
This work was supported in part by the National Natural Science
Foundation of China under Contract No.10675167.
\end{acknowledgments}

%%%%%%%%%%%%%%%%%%%%%%%%%%%%%%%%%%%%%%%%%%%%%%%%%%%%


\begin{thebibliography}{99}

\bibitem{sigma-b}T.~Aaltonen {\it et al.} (CDF Collaboration), Phys. Rev. Lett. {\bf 99}, 202001 (2007).

\bibitem{splitting}S.~B.~Athar {\it et al.} (CLEO Collaboration), Phys. Rev. D {\bf 71}, 051101 (2005).

\bibitem{lamdarc}B.~Aubert {\it et al.} (BABAR Collaboration), Phys. Rev. D {\bf 72}, 052006 (2005).

\bibitem{lamdar-b}D.~Acosta {\it et al.} (CDF Collaboration), Phys. Rev. Lett. {\bf 96}, 202001 (2006).

\bibitem{quark model}D.~Ebert, R.~N.~Faustov, and V.~O.~Galkin, Phys. Rev. D {\bf 72}, 034026 (2005).

\bibitem{quark model 2}S.~Capstick and N.~Isgur, Phys. Rev. D {\bf 34}, 2809 (1986).

\bibitem{mass formular}R.~Roncaglia, D.~B.~Lichtenberg, and E.~Predazzi, Phys. Rev. D {\bf 52}, 1722 (1995).

\bibitem{lattice}N.~Mathur, R.~Lewis, and R.~M.~Woloshyn, Phys. Rev. D {\bf 66}, 014502 (2002).

\bibitem{svzsum}M.~A.~Shifman, A.~I.~Vainshtein and V.~I.~Zakharov, Nucl. Phys. {\bf B147}, 385 (1979); {\bf B147}, 448 (1979);
 V.~A.~Novikov,M.~A.~Shifman, A.~I.~Vainshtein and V.~I.~Zakharov, Fortschr. Phys. {\bf 32}, 585 (1984).

\bibitem{evs}E.~V.~shuryak, Nucl. Phys. {\bf B198}, 83 (1982).

\bibitem{alfa}A.~G.~Grozin and O.~I.~Yakovlev, Phys. Lett. B {\bf 285}, 254 (1992); {\bf 291}, 441 (1992);
S.~Groote, J.~G.~K\"{o}rner, and O.~I.~Yakovlev, Phys. Rev. D {\bf55},
3016 (1997).

\bibitem{dai}Y.~B.~Dai, C.~S.~Huang, C.~Liu, C.~D.~L\"{u}, Phys. Lett. B {\bf371}, 99 (1996).

\bibitem{mqhuang}D.~W.~Wang, M.~Q.~Huang, and C.~Z.~Li,
Phys. Rev. D {\bf 65}, 094036 (2002); D.~W.~Wang, and M.~Q.~Huang,
Phys. Rev. D {\bf 67}, 074025 (2003).


\bibitem{EBagan}E.~Bagan, M.~Chabab, H.~G.~Dosch, and S.~Narison, Phys. Lett. B
{\bf 278}, 367 (1992); {\bf 287}, 176 (1992).

\bibitem{ksi}Francisco O.~Dur\~{a}es, Marina Nielsen, Phys. Lett. B
{\bf 658}, 40 (2007).

\bibitem{zhigang}Z.~G.~Wang, Eur. Phys. J. C. {\bf 54}, 231 (2008).

\bibitem{new}M.~Karliner, B.~K.~Zur, H.~J.~Lipkin, and J.~L.~Rosner,
arXiv:0708.4027; X.~Liu, H.~X.~Chen, Y.~R.~Liu, A.~Hosaka, and
S.~L.~Zhu, Phys. Rev. D {\bf 77}, 014031 (2008); W.~Roberts and
M.~Pervin, arXiv:0711.2492.

\bibitem{kim}Hungchong Kim, Su Houng Lee, Yongseok Oh, Phys. Lett. B
{\bf 595}, 293 (2004).

\bibitem{bracco}M.~E.~Bracco, A.~Lozea, R.~D.~Matheus, F.~S.~Navarra and M.~Nielsen, Phys. Lett. B
{\bf 624}, 217 (2005); M.~Nielsen, R.~D.~Matheus, F.~S.~Navarra,
M.~E.~Bracco, and A.~Lozea, Nucl. Phys. B, Proc. Suppl. {\bf 161},
193 (2006).

\bibitem{Ioffe}B.~L.~Ioffe, Nucl. Phys. {\bf B188}, 317 (1981).

\bibitem{cohen}T.~D.~Cohen, R.~J.~Furnstahl, D.~K.~Griegel, and
X.~M.~Jin, Prog. Part. Nucl. Phys. {\bf 35}, 221 (1995).



\bibitem{reinders}L.~J.~Reinders, H.~Rubinstein and S.~Yazaki, Phys. Rep. {\bf 127}, 1
(1985).

\bibitem{PDG}W.~M.~Yao {\it et al.} (Particle Data Group), J. Phy. G {\bf
33}, 1 (2006).



\bibitem{borel}F.~S.~Navarra, M.~Nielsen, and S.~H.~Lee, Phys. Lett. B
{\bf 649}, 166 (2007); R.~D.~Matheus, F.~S.~Navarra, M.~Nielsen, and
R.~Rodrigues da Silva, Phys. Rev. D {\bf 76}, 056005 (2007).

\bibitem{Narison}S. Narison, Phys. Lett. B {\bf  605}, 319 (2005).
\end{thebibliography}
\end{document}